\documentclass{osa-article}

\journal{ome}
\articletype{Research Article}
\usepackage{amsfonts,amsmath} 
\usepackage{graphicx,epsfig,psfrag}% Include figure files 
\usepackage{color}
\usepackage{url}
\usepackage[breaklinks=true]{hyperref}
\usepackage{mathtools}
\usepackage{subfigure}
\hypersetup{
        colorlinks = true,
        citecolor = blue
}
\usepackage{braket}
\usepackage{xcolor}

\def \lod{\lambda^{\text{od}}}
\def \ld{\lambda^{\text{d}}}
\def \Vod{V^{\text{od}}}
\def \Vd{V^{\text{d}}}
\def \vSl{v^\text{S}_l}
\def \VS{V_{\text{IAAF}}}

\begin{document}

\title{Topology in quasicrystals}

\author{Oded Zilberberg\authormark{*}}

\address{Institute  for  Theoretical  Physics,  ETH  Zurich,  8093  Z{\"u}rich,  Switzerland}

\email{\authormark{*}odedz@phys.ethz.ch} %% email address is required
\homepage{https://quest.phys.ethz.ch/} %% author's URL, if desired


\begin{abstract}
Topological phases of matter have sparked an immense amount of activity in recent decades. Topological materials are  classified by topological invariants that act as a non-local order parameter for any symmetry and condition. As a result they exhibit quantized bulk and boundary observable phenomena,  motivating various applications that are robust to perturbations.  In this review, we explore such a topological classification for quasiperiodic systems, and detail recent experimental activity in the field.
\end{abstract}

%%%%%%%%%%%%%%%%%%%%%%%%%%  body  %%%%%%%%%%%%%%%%%%%%%%%%%%
\section{Introduction}

The realization that solid structures form mostly as periodic crystalline structures allowed us to describe numerous material properties in nature~\cite{Ashcroft76}. Specifically, as there is only a finite number of possible periodic arrangements in each dimension, we can categorize the material spectral properties (band structures) according to their spatial symmetries.  Using this general framework, we obtain an overview of material bulk responses. 
In recent decades, concepts originating from the mathematical field of topology bear growing impact on our understanding of material properties~\cite{bernevig2013topological,ozawa2019topological}. Topology generalizes the notion of symmetry by attributing non-local order parameters (topological invariants) to the material, which are used to classify materials into distinct families. These families cannot be continuously deformed into one another, implying the existence of observable topological phase transitions. 

In many cases, the topological invariant is also observable through quantized bulk responses  with associated boundary phenomena at the open end of the material. The most known example of such a response is the quantum Hall effect: a 2D electron gas in the presence of a perpendicular magnetic field exhibits a quantized transverse (Hall) conductance, where the quantization originates from topological invariants (Chern numbers) associated with the system's bandstructure~\cite{klitzing1980new, TKNN,cage2012quantum}. In correspondence to the quantized bulk response, chiral edge modes appear at the boundary of the system~\cite{halperin1982quantized, macdonald1984quantized,streda1987edge, bernevig2013topological,ozawa2019topological,frohlich2018chiral}. Note that the non-locality of the topological invariants implies a robustness to perturbations such as disorder - a fact that promoted the quantum Hall effect as a standard for metrology~\cite{von2019essay}. 

The topology in the quantum Hall effect originates from a length scale competition between the magnetic field and the translation invariance of the electron's kinetic energy~\cite{Harper:1955,TKNN}. When the electrons hop on a lattice, this competition can also lead to incommensurate situations, where the system is no longer invariant to translations~\cite{Hofstadter}. Such is also the case in quasicrystals, which are ordered but non-periodic solid structures~\cite{senechal_quasicrystals_1995,lifshitz2003quasicrystals}. The long-range order of quasicrystals can be projected from periodic higher-dimensional systems~\cite{bohr1925theorie,bohr1926theorie,penrose1974role}, and thus exhibits sharp diffraction peaks~\cite{shechtman1984metallic,levine1984quasicrystals,yamamoto1996crystallography}. Spectral characteristics of such quasiperiodic patterns have been studied in a variety of platforms, revealing highly fractal energy
levels with infinitely many energy gaps, self-similar eigenstates and critical localization~\cite{aubry1980,kohmoto1983,kohmoto_cantor_1984,Kohmoto1986,Kohmoto1987,tanese2014fractal}. Interestingly, the topological characterization of quasicrystals can also be mapped to higher-dimensions using a mapping to the quantum Hall effect, and associating bulk gaps with Chern numbers~\cite{Kraus2012a,Kraus2013,Kraus2016,prodan2015virtual,prodan2016bulk}. This mapping complements earlier topological classifications of quasiperiodic systems by gap labeling approach~\cite{bellissard1989spectral,bellissard_continuity_1991,bellissard_gap_1992}. 

Here, we review the mapping between quasiperiodic models and quantum Hall systems. In section~\ref{sec:QCs}, we introduce the concept of quasicrystals and their corresponding quasiperiodic models. Section~\ref{sec:dimext} is devoted to the relation of some families of quasiperiodic models to the quantum Hall effect, whereas in section~\ref{sec:topPump}, we present the proof from Ref.~\cite{Kraus2012a} on how to associate Chern numbers to gaps of the quasiperiodic models. In section~\ref{sec:opt}, we then shortly review recent experiments in various optical setups that explore the topological properties of quasicrystals.

\section{Quasiperiodic order and approximants}
\label{sec:QCs}
\begin{figure}[t!]
\centering\includegraphics[width=\textwidth]{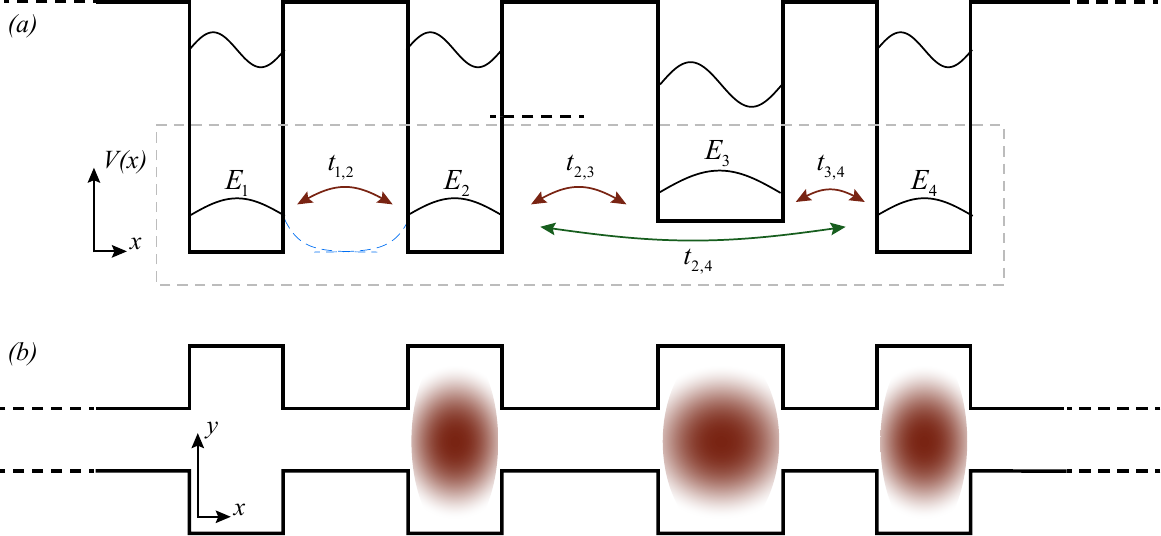}
\caption{A segment in a quasiperiodic arrangement of potential wells. (a) A 1D quasiperiodic arrangement of potential wells. Each well has a specific height and width, and corresponding bound modes. A quasiperiodic tight-binding limit model is constructed, for example, from the lowest state in each well $n$ with energy $E_n$. Through evanescent wave overlap of the states (dashed blue line example), we obtain variable-range hopping amplitudes $t_{i,j}^{\phantom{\dagger}}$, e.g., shown are nearest-neighbor (red arrows) and next-nearest-neighbor hopping (green arrow). The quasiperiodicity in the potential shape translates to a quasiperiodic modulation in the tight-binding model, cf.~Eq.~\eqref{Eq:HS_1D}. (b) A similar construction of a quasiperiodic 1D chain for waves moving in a sculptured 2D medium, where the lowest bound modes of each box are shown (red gradients). Such a construction readily extends to other quasiperiodic arrangements in 2D and 3D. }\label{fig1}
\end{figure}

Quasicrystals are materials with atoms arranged in a non-periodic but ordered pattern. To describe their electronic properties, we consider hybridization of their valence electrons similar to what is done for periodic materials, e.g., by constructing a Kronig–Penney model for nearly-free electrons in the corresponding quasiperiodic potential energy landscape~\cite{Ashcroft76}, see Fig.~\ref{fig1} for an illustration of a quasiperiodic 1D chain. As we discuss below, such a construction lends itself also to metamaterial structures, where waves interfere in the presence of a patterned medium. Assuming sufficiently-spaced and sufficiently-deep potential wells allows labeling of the resulting spectral band structures according to the separated orbits of the individual potential wells, e.g., by s-, p-, and d-like orbital bands. On top of this orbital separation, the longer-ranged quasiperiodic spatial arrangement of the ``atoms'' leads to splitting of the bands into infinitely-many mini-bands.

Considering, for simplicity, the resulting tight-binding Hamiltonian description of the s-orbits hybridization along a 1D quasiperiodic chain, we have
\begin{align}
 \label{Eq:HS_1D}
    H_{s} = \sum_n \left\{\ld \Vd_{n} c_n^\dag c^{\phantom{\dag}}_n + \sum_{m >0}\left[ \left(t^{\phantom{\dag}}_{m}+\lod_{m} \Vod_{n,m}\right)
    c_n^\dag c^{\phantom{\dag}}_{n + m} + h.c.\right] \right\}\,,
\end{align} 
where $n,m$ are integers iterating over the atomic sites along the chain, $c_n^\dag$ is a second-quantized creation operator at site $n$, $\Vd_{n}$ is an on-site quasiperiodic potential corresponding to the height of the potential-well at each site with modulation amplitude $\ld$. The kinetic energy matrix elements are decomposed into  hopping amplitudes $t_m$ associated with the mean hybridization between orbits separated by $m$ sites, on top of which a quasiperiodic function $\Vod_{n,m}$ of amplitude $\lod_{m}$ encodes the modulated inter-atomic distances. Note that generally we can also consider additional quasiperiodic potentials, e.g., spin-dependent and spin-orbit-dependent terms.

\begin{figure}[t!]
\centering\includegraphics[width=\textwidth]{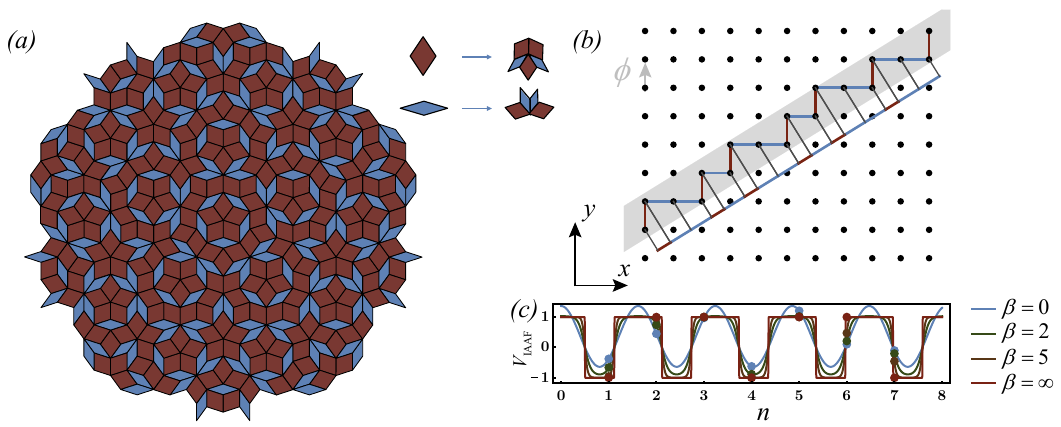}
\caption{Methods to generate quasiperiodic tiling and potentials. (a) Substitution rules for rhombic tiles (upper right) are applied to create a 2D Penrose tiling. (b) The cut-and project method for generating 1D quasiperiodic chains. A cut strip (gray) is tilted in an angle with tangent $\tau$ relative to a 2D square lattice. The lattice points within the cut's acceptance window are projected onto a 1D line, with two possible spacings (red and blue). (c) Sampling of the Interpolating Aubry-Andr\'e-Fibonacci potential at integer values, cf.~Eq.~\eqref{pot}. Periodicity with respect to $\phi$ in (c) is mappable to periodicity with respect to displacement of the cut strip in (b)~\cite{kraus2012}.  }\label{fig2}
\end{figure}

The quasiperiodic 1D functions $\Vd_{n}$ and $\Vod_{n,m}$ can be constructed, for example, by primitive substitutions~\cite{senechal_quasicrystals_1995, fogg_substitutions_2002} to iteratively generate a non-repeating sequence of discrete values. Known examples involving two discrete values $a$ and $b$ are the Fibonacci ($a\rightarrow ab$, $b\rightarrow a$)~\cite{kohmoto1983, ostlund1983,levine1984quasicrystals}, Thue-Morse ($a\rightarrow ab$, $b\rightarrow ba$)~\cite{bellissard_gap_1992} and Period-doubling sequences ($a\rightarrow ab$, $b\rightarrow aa$)~\cite{bellissard_gap_1992,damanik1998singular}, see Fig.~\ref{fig2}(a) for a 2D example. Alternatively, the method of ``cut-and-project'' [see Fig.~\ref{fig2}(b)] can be employed to generate step-wise (Sturm) functions of the form $V_{\rm Sturm}(x)=2( \lfloor (x+2)/\tau \rfloor - \lfloor (x+1)/\tau \rfloor ) - 1 = \pm1$, where $\tau$ is an irrational number,  $\lfloor x \rfloor$ is the floor function, and the potentials in Eq.~\eqref{Eq:HS_1D} sample the continuous potential at integer values $V(n)$, thus, enforcing the incommensuration between the periodic lattice and the potential modulation lengthscales~\cite{senechal_quasicrystals_1995,kraus2012}. When $\tau = (1 + \sqrt{5})/2$ is
the golden ratio, this method also yields the Fibonacci sequence. As another example, quasiperiodic sequences with infinitely-many values can be obtained by an incommensurate sampling of a periodic function, e.g., $f(n/\tau +\phi)$, which is periodic under a scan of $\phi$ from 0 to $2\pi$, see Fig.~\ref{fig2}(c). A useful example for our discussion is the interpolating Aubry-Andr\'{e}-Fibonacci (IAAF) function~\cite{kraus2012, Verbin2013, Verbin2015}
\begin{equation} \label{pot}
	\VS (\beta,\tau,x, \phi)=	-\frac {\tanh {\left[ \beta (\cos{(2\pi x/\tau + \phi)} -\cos{(\pi/\tau)}) \right]}} {\tanh \beta}\,,
	\end{equation}
where $\phi$ encodes a `phasonic' continuous shift of the potential, and $\beta$ interpolates between two known limiting cases, namely, for $\beta \rightarrow 0$, we obtain the Harper-Aubry-Andr\'e cosine modulation~\cite{harper1955,aubry1980}, up to a constant shift $\left[\cos{(2\pi x/\tau + \phi)} -\cos{(\pi/\tau)}\right]$, and for $\beta \rightarrow \infty$ the  Fibonacci sequence is reached with alternating values of $\pm 1$~\cite{kohmoto1983, ostlund1983,kraus2012}. Note that in dimensions $D>1$ the quasiperiodic potential functions become more complex and include also rotational symmetries that are forbidden for periodic crystals~\cite{senechal_quasicrystals_1995}, see Fig.~\ref{fig2}(a). 

From Eq.~\eqref{Eq:HS_1D}, we can identify several branching points associated with how different fields analyze quasicrystals.  For example, we can either (i) study directly the ab-initio atomic arrangement and hybridization as discussed above, (ii)  analyze the most general spectral properties that can be obtained from a specific quasiperiodic atomic arrangement, i.e., without turning to ab-initio extraction of parameters; this is similar to symmetry group analysis of spectral features of periodic crystals, and (iii) analyze specific sub-cases of the general case~\eqref{Eq:HS_1D}, e.g., by considering solely short-ranged potentials. In the spirit of (ii), much research is pursued within the mathematical physics community, aiming to answer (a) whether the spectrum has a well-defined structure with a countable number of bands and gaps [see, e.g., Ref.~\cite{bellissard_gap_1992}], (b) whether the eigenstates of the quasicrystals are spatially-extended or localized [see, e.g., Refs.~\cite{aubry1980,jitomirskaya1999,strkalj2020}], and (c) whether a topological classification of spectral bands of the quasicrystals is possible by harnessing K-theoretical tools [see e.g., Refs.~\cite{bellissard_gap_1992,kellendonk2019bulk} and references therein]. Note that (a) and (c) are often combined, as specific quasiperiodic sequences fix the integrated density of states as a topological index that allows for labeling of spectral gaps of the material, and thus their enumeration~\cite{bellissard_gap_1992}.  In the context of (iii), different approximations are taken, such as assuming only short-ranged nearest-neighbor hopping ($t^{\phantom{\dag}}_{m\neq 1}=\lod_{m\neq 1}=0$), on top of which only \textit{diagonal} ($t^{\phantom{\dag}}_{1}=\lod_{1}=0$) or \textit{off-diagonal} ($\ld=0$) models are explored. Note that the large variety of such approximations in dimensions larger than 1 leads to different models exhibiting different effects, while describing the same long-ranged quasiperiodic spatial modulation.

Regardless of the approach or approximation discussed above, the long-range nature of the quasiperiodic potentials does not allow us to apply Bloch's theorem to obtain quasimomenta as good quantum numbers and thus simplify the solution of the problem. Similarly, we cannot solve exactly the spectrum of the system on a computer as the quasiperiodicity manifests as a deviation from periodicity at all length scales, i.e., a true quasiperiodic system is infinitely large. Similar challenges also occur in the study of disordered systems. There the thermodynamic properties of the disordered material are approached by studying the impact of the disorder on a finite system and observing how the effect scales with the system's size. Crucially, a disorder-average is taken in the analysis, such that the general (noise) ensemble properties are explored, and not specific (sample path) cases. In comparison, for quasiperiodic systems, we can also study the spectral properties of finite-sized samples from the long-ranged system, and average over them, with a reduced statistical averaging arising from the self-similar nature of quasicrystalline eigenstates, see, e.g., Ref.~\cite{strkalj2020} and references therein. 

\begin{figure}[t!]
\centering\includegraphics[width=0.8\textwidth]{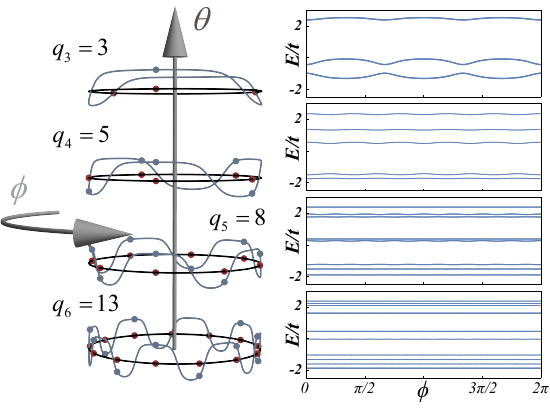}
\caption{Approximant path and its spectrum for the 1D diagonal model~\eqref{Eq:HS_1D} with an on-site quasiperiodic potential~\eqref{pot} at $\beta=2$ and approaching $\tau=(1+\sqrt{5})/2$. (left) The approximant model is placed on a ring with periodic boundary conditions. Red dots mark atomic sites, which sample the on-site potential (blue line) at the blue dots. Shown are the approximants with unit cells of length $q_n=\{3,5,8,13\}$. The bulk spectrum of the approximant is spanned by the phase twist $\theta$ at the periodic boundary, which is equivalent to threading a flux through the ring. The potential~\eqref{pot} can be displaced as a function of the phason $\phi$. (right) The corresponding spectrum of the approximants with potential amplitude $\ld/t_1=0.9$ as a function of $\phi$ for $\theta=0$. We observe splitting of the states into mini-bands, where large features (gaps) are clearly robust along the approximant path while new (smaller) features appear. }\label{fig3}
\end{figure}

We can study the bulk spectral properties of the quasiperiodic model along a so-called \textit{approximant path}, where we explore finite periodic models that capture well the incommensurate length scales competition within their enlarged unit cell~\cite{senechal_quasicrystals_1995,beckus_spectral_2018}. Beyond the finite unit cell length scale, we use a periodic approximation and apply Bloch’s theorem. We recall that when the approximant length scale in real space is equivalent to $q$, the corresponding first Brillouin zone in momentum space is $q$ times smaller than for a simple periodic lattice. As a consequence, the energy dispersion $E(k)$ of the periodic chain folds from the larger to the reduced 1st Brillouin zone, thus forming $q$ different energy bands. Note that the bands may overlap depending on the specific model, e.g., if a symmetry persists under the folding. As an example for such an approximant approach, we consider the interpolating quasiperiodic function~\eqref{pot}, see Fig.~\ref{fig3}. The irrational modulation $\tau$ has a unique continued fraction representation
\begin{align}
    \frac{1}{\tau}=a_0 + \cfrac{1}{a_1 + \cfrac{1}{a_2 + \cfrac{1}{ \ddots + \cfrac{1}{a_n} }}}\,,
    \end{align}
where $a_i$ are integer numbers. Thus, while scaling the system size, at the $n^{\rm th}$ iteration, we can take a rational modulation $\tau_n=q_n/p_n$, that leads to a periodic chain with $q_n>q_{n-1}$ bulk bands. Clearly, the difference between the $n-1$ and $n$ iteration occurs at relatively longer length scales, implying that the differences between the bulk bands of different approximants will occur on small energy scales, i.e., by splitting of bulk bands into smaller bands. For the case of the Fibonacci sequence, the approximant path grows as the Fibonacci numbers with $a_0=0$ and $a_{i\neq 0}=1$, corresponding to denominators $q_n\in\{1,2,3,5,8,13,21,34,\ldots\}$. Approximant paths exist also for other quasiperiodic orders, as well as in higher-dimensional quasicrystals, where the most important challenge is to identify properties that survive the scaling to the thermodynamic limit. 

A prominent topological classification approach for quasicrystals involves the labeling of spectral gaps that are robust along the approximant path. Once a spectral gap appears, at sufficiently later approximants along the scaling, the formation of small mini-bands cannot close the robust (larger) energy gap, and thus the integrated density of states remains constant under the scaling and acts as a topological index, which cannot be changed unless a perturbation is introduced that closes that gap.  

\section{Dimensional extension and relation to the quantum Hall effect}
\label{sec:dimext}
\begin{figure}[t!]
\centering\includegraphics[width=\textwidth]{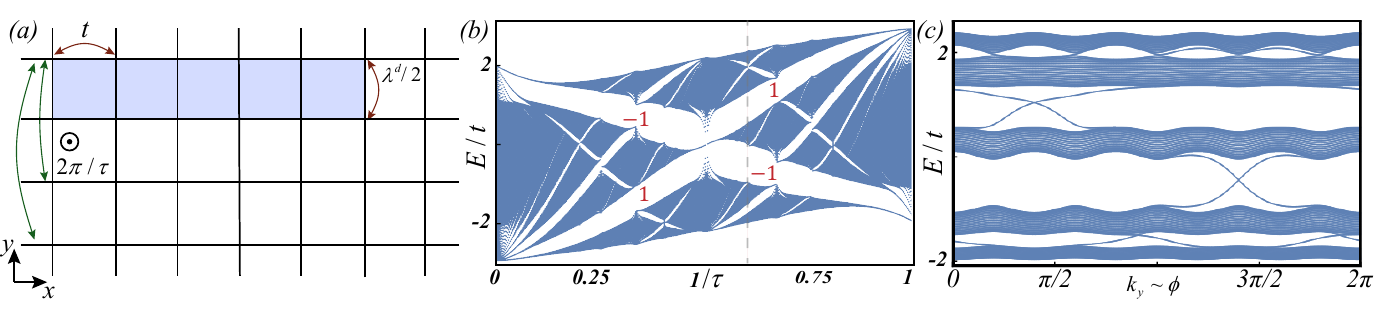}
\caption{Two-dimensional quantum Hall effect on a lattice. (a) Illustration of the 2D model obtained by dimensional extension~\eqref{Eq:HS_2D} with nearest-neighbor hopping along the $x$- and $y$-directions, as well as longer-ranged hopping along $y$ depending on $\beta$. A magnetic flux $2\pi/\tau$ pierces each square plaquette, leading to an enlarged magnetic unit cell (blue box for $q_n=5$). (b) Hofstadter butterfly spectrum as a function of $1/\tau$ for the diagonal model with $\beta=2$ and $\ld/t=0.9$. The Chern number of the large gaps is marked (red), cf.~Eq.~\eqref{Eq:Chern}. (c) The spectrum of the same system for $\tau=5/3$ [marked with gray dashed line in (b)] with open boundary conditions, exhibiting 5 bands and 4 gaps. Topological edge modes cross the bulk gaps in accordance with the bulk-boundary correspondence. As the quasimomentum $k_y$ of the quantum Hall system~\eqref{Eq:HS_2D} is equivalent to the pump parameter $\phi$, we observe the boundary phenomena of topological pumps, namely, at the 0D boundary of the 1D pump, boundary modes will appear and thread across the gap.  }\label{fig4}
\end{figure}

We now focus on the specific family of models including only an on-site (diagonal)  interpolating quasiperiodic potential~\eqref{pot}, and study their dependence on the parameter $\phi$.
For any given $\phi$,
we can understand the potential to be originating from a quasimomentum in some perpendicular (non-physical) direction. In other words, the combined real space of the chain with the compact parameter $\phi$ is two-dimensional. It is illuminating to extend our description to this additional perpendicular axis by  performing an inverse Fourier map of the model with respect to the parameter $\phi$. We thus obtain a 2D real space \textit{ancestor model}~\cite{kraus2012}
\begin{align} \label{Eq:HS_2D}
    H_{\rm 2D} =\sum_{n,m}& \Big[ t c_{n,m}^\dag c_{n + 1,m}  + \ld\sum_{l=0}^\infty \vSl(\beta) e^{i\pi (2n+3)l/\tau} c_{n,m}^\dag c_{n,m+l}  + h.c. \Big]\,, 
\end{align}
where $\vSl(\beta) = \int_0^{2\pi} (d\phi/2\pi) e^{il \phi} \, \VS(\beta,\tau,x, \phi)$. The  model describes particles hopping on a 2D square lattice in the presence of a perpendicular magnetic field with  $2\pi/\tau$ flux threading each square plaquette written in the Landau gauge using Peierls substitution~\cite{peierls1933theorie,luttinger1951effect}, see Fig.~\ref{fig4}(a). Along the $x$-axis, we have nearest-neighbor hopping as in the descendent 1D model. Along the $y$-axis, we have long-ranged hopping depending on $\beta$, where in the limit of $\beta\rightarrow 0$, it also becomes short-ranged and we obtain the 2D model Hofstader model~\cite{Harper:1955,azbel1964energy,Hofstadter}
\begin{equation} \label{eq:HdH_2D}
    H_{\rm Hof} = \sum_{n,m} \Big[ t c_{n,m}^\dag c_{n + 1,m}
    + \frac{\ld}{2} e^{i2\pi n/\tau} c_{n,m}^\dag c_{n,m + 1} + h.c. \Big]\,. %\nonumber
\end{equation}

Interestingly, we now mapped the length scales competition between the 1D lattice and the quasiperiodic potential in Eq.~\eqref{Eq:HS_1D} to a competition between the 2D lattice and the magnetic length scale in Eqs.~\eqref{Eq:HS_2D} and~\eqref{eq:HdH_2D}. In the absence of the magnetic field, $\tau\rightarrow\infty$, the Hamiltonian~\eqref{eq:HdH_2D} commutes with the group of lattice translations, whereas when the magnetic field breaks this
symmetry, the notion of the magnetic translation group was introduced~\cite{Zak}. It is generated by
the operators $T_{\hat{m}}$ and $T_{\hat{n}}$, where $T_{\hat{m}} c_{n,m} T_{\hat{m}}^{-1} = c_{n,m+1}$ and $T_{\hat{n}} c_{n,m}
T_{\hat{n}}^{-1} = e^{-i2\pi m/\tau} c_{n+1,m}$. For an irrational $\tau$, these operators commute with the Hamiltonian but not with each other. Nevertheless, similar to the approximant discussion above, for a rational flux, $\tau=q/p$, the operator $T_{q\hat{n}} = (T_{\hat{n}})^q$ commutes with $T_{\hat{m}}$. We can therefore 
simultaneously diagonalize $H_{\rm 2D}$, $T_{q\hat{n}}$, and $T_{\hat{m}}$. Here, too, the enlarged real space magnetic unit cell implies folding of the corresponding quasimometa (now onto a 2D torus) and the formation of $q$ bands in the spectrum, thus generating Hofstadter butterfly spectra, see Fig.~\ref{fig4}(b)~\cite{Hofstadter}. 

Using the 2D vantage point, the magnetic length scale competition allows for another topological characterization of the system. Specifically, each gap $r$ in the spectrum of this
model is associated with a quantized and nontrivial Chern number~\cite{TKNN,streda1982theory,avron1983homotopy, kohmoto1985topological, dana1985quantised}
\begin{align} \label{Eq:Chern}
\nu_r & =\frac{1}{2\pi i}\int_{0}^{2\pi}d\phi d\theta\, C_r(\theta,\phi)\,,
\end{align}
 where $\phi$ and $\theta$ are phase twists (fluxes) associated with the periodic boundary conditions along the $y$- and $x$-directions, and  
\begin{align}
C_r(\phi,\theta) & =\textrm{Tr}\left\{P_r\left[ \partial_\phi P_r,\partial_\theta P_r\right]\right)\,,\label{Eq:Chern-1}
\end{align}
is the Chern density (curvature) with $[a,b]$ the commutator of $a$ and $b$ and the projector on the states below the chosen spectral gap
\begin{align}
P_r(\phi,\theta) & =\sum_{E_{n}<E_\text{gap}}|n\rangle\langle n|\,.
\label{projector}
\end{align}
Here, $|n\rangle$ is an eigenstate of $H_{\rm 2D}$ with energy $E_n$, and $E_\text{gap}$ is
the energy at the center of the gap. For a sufficiently large system size, such that the gaps are open for any $\phi$ and $\theta$, $P_r(\phi,\theta)$ is a well-defined quantity, see Fig.~\ref{fig4}(b).
Importantly, this topological characterization manifests as a quantized  bulk response of the system, namely, quantized Hall conductance~\cite{klitzing1980new,laughlin1981quantized, TKNN, niu1985quantized}. Furthermore, the quantized bulk response also implies the appearance of chiral edge modes that cross the spectral gap and are localized at the boundary of the system, known as the bulk-boundary correspondence, see Fig.~\ref{fig4}(c)~\cite{halperin1982quantized, macdonald1984quantized,streda1987edge, bernevig2013topological,ozawa2019topological,frohlich2018chiral}.  

Recall that for any rational $\tau=q/p$, we have $q$ bands and maximally $q-1$ gaps in the spectrum, cf. discussion on approximants above and Fig.~\ref{fig4}(b). Due to the symmetry properties of the magnetic translation group, there is an interesting connection between number theory and the Chern number $\nu_r$ 
associated with a gap number $r=1,\ldots,(q-1)$, namely it satisfies the Diophantine equation $r = \nu_r p + t_r q$, where $\nu_r$ and $t_r$
are integers~\cite{TKNN,streda1982theory,dana1985quantised,avron1986generic}. This condition has infinitely many possible solutions, as $\nu_r$ is only well defined modulo $q$. For specific models, the Chern numbers are constrained to provide a single solution for the Diophantine equation, e.g., for the Hofstadter model, where $0 < |\nu_r| < q/2$~\cite{TKNN}. For other models, finding the correct Chern numbers requires evaluation of Eq.~\eqref{Eq:Chern} and is called ``coloring Hofstadter butterflies'' due to the association of a color to each value of the Chern number  in a spectral gap~\cite{avron2014study,agazzi2014colored}. Crucial for our discussion, we note also that we can divide the Diophantine equation by $q$ to obtain $\rho_r = \nu_r/\tau  + t_r$, which defines the Chern number for any particle density (filling factor) $\rho_r$ within a spectral gap. The latter result does not rely on approximants, it can be applied also for quasiperiodic models, and is sometimes referred to as Streda’s formula~\cite{streda1982theory, bernevig2013topological}. Note that the filling factor (a.k.a.~integrated density of state) $\rho$ is fixed for quasiperiodic models in each gap by the above condition, i.e., it cannot change without a change in $\tau$. 

\begin{figure}[t!]
\centering\includegraphics[width=\textwidth]{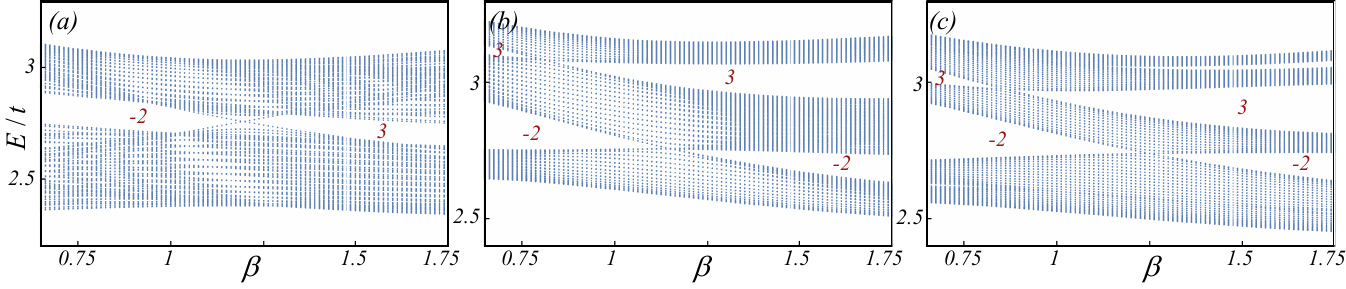}
\caption{Phase transitions along the approximant path and interpolation deformation~\eqref{pot}. Zoom in on the upper bands in the spectrum of the diagonal IAAF, as a function of $\beta$ for approximants (a) $\tau=5/3$, (b) $\tau=8/5$, and (c) $\tau=13/8$. Gap closings can occur as a function $\beta$, but a change of the Chern number (red labels) is only allowed in specific cases when it complies with the Diophantine equation. }\label{fig5}
\end{figure}

Following the approximant path and moving back towards quasiperiodic models, we approach an irrational magnetic flux  by taking an appropriate rational limit with $p,q \rightarrow \infty$. In this limit, the
spectrum becomes a Cantor set fractal~\cite{Hofstadter}. At each approximant generation, new gaps can open, exhibiting quantized Chern numbers. It is important to note that depending on the model, the gaps can close and reopen under some perturbation that does not break the unit cell, see Fig.~\ref{fig5}. In such a case, the Diophantine equation still holds and the corresponding Chern numbers can change by a multiple of $q$ modulo $q$. Nevertheless, at later approximants, due to the changed denominator $q_{n+l}> q_n$, such a exchange of Chern numbers is no longer possible and the gap may close but without a change to the Chern number. Crucially, in order to change a topological index of the gap, a phase transition must occur and the gap must close, i.e., there is no way to deform between the two distinct topological states without the phase transition. However, picking a specific deformation [as for example done in the IAAF model~\eqref{pot}], gaps can close and reopen, but it does not necessitate a change in the topological index. Combining the two arguments in this paragraph, alongside the fact that all gaps are known to be open in both limits of the IAAF defomration, we could color the gaps of the Fibonacci limit of the model~\cite{kraus2012}.

We can employ dimensional extension of quasiperiodic models in $D>1$ with spatial modulations along orthogonal axes to obtain quantum Hall effects in $2\times D>2$, which are then characterized by higher dimensional Chern numbers, families of associated bulk responses, as well as, associated boundary effects with co-dimension $d\in\{1,\ldots,D\}$~\cite{Kraus2013,Kraus2016,price2015four, price2016measurement,ozawa2016synthetic,zilberberg2018photonic,lohse2018exploring,petrides2018six,petrides2020higher}. For such cases, similar approximant paths apply with generalized translation operators as discussed above. At the same time, it becomes much harder to identify gaps that remain open, and the gap labeling involves bookkeeping that tracks the labels from the orthogonal axes. Furthermore, introducing terms that mix the orthogonal spatial axes, as is expected for more general quasicrystals [cf.~Fig.~\ref{fig2}(a)], implies that a mapping to a $2\times D$ ancestor model may be possible [see, e.g., recent work with hexagonal moir\'e structures~\cite{rosa2020topological}], but the gap labeling approach detailed here will no longer hold.

\section{Topological pumps, dimensional reduction, and Chern numbers\\ of the quasiperiodic model}
\label{sec:topPump}
We have seen a relationship between the family of 1D models~\eqref{Eq:HS_1D} parametrized by $\phi$ and the 2D quantum Hall effect~\eqref{Eq:HS_2D}. Such a relationship allows for the prediction of quantized charge transport in temporally-modulated chains, i.e., topological charge pumping~\cite{laughlin1981quantized,Thouless1983,Kraus2012a, lohse2016thouless}. Here, the time-dependent and adiabatic threading over the parameter $\phi$ allows for the 1D family of models to cover the 2D space spanned by both $\phi$ and $\theta$, cf.~Eq.~\eqref{Eq:Chern}. This, in similitude to the quantized Hall conductance, leads to a quantized number of charges being pushed across the chain per cycle of $\phi$ from 0 to $2\pi$, with the quantization originating from the Chern number of the gap. Note that such a topological bulk response is associated with the whole pump cycle, and implies a specific boundary phenomena: at the 0D boundary of the pump, subgap modes cross the topological gap to carry the quantized charge across the system~\cite{halperin1982quantized,Kraus2012a,hatsugai2016bulk,ozawa2019topological}. This can be thought of as an iteration with $\phi$ over the dispersive and chiral 1D edge modes of the 2D quantum Hall effect, see Fig.~\ref{fig4}(c).  

Similar to the 2D quantum Hall effect on a lattice, the topological pump is characterized by a 2D Chern number and exhibits quantized charge transport for both rational and irrational $\tau$. We, now, focus instead on the direct topological characterization of the quasiperiodic chain using Chern numbers, i.e., without the need to use the parameter $\phi$, where we summarize and repeat the proof shown in Ref.~\cite{Kraus2012a}, cf.~also Refs.~\cite{prodan2015virtual,bourne2018chern}. To do so, we employ the approximant path described above, and place at each iteration the chain on a ring (periodic boundary conditions) of length $q_n$. We also introduce a flux $\theta$ (phase twist) threading the ring, cf.~Fig.~\ref{fig3}.  The periodic model has the same spectral gaps as those of the original chain up to small corrections~\cite{niu1985quantized}, whereas in the thermodynamic limit the effect of $\theta$
completely vanishes~\cite{niu1985quantized,ringel2011determining}.

We now show that the impact of $\phi$ is also vanishing for quasiperiodic chains. Consider the impact of a shift of the parameter $\phi \rightarrow \phi + \epsilon$, where $\epsilon=2\pi l/q_n$ and
$l=0,1,...,L-1$. Under this choice of $\epsilon$, the magnetic translation symmetry discussed above implies that there is always a spatial translation of $n
\rightarrow n + n_\epsilon$ lattice sites, where $n_{\epsilon}\in0,1,...,q_n-1$, such that the potential~\eqref{pot}, has $V_{\rm IAAF}(x,\phi+\epsilon)=V_{\rm IAAF}(x+n_\epsilon,\phi)$.  Therefore, for each shift $\epsilon$ of $\phi$, we can associate a spatial translation operator $\hat{T}_{\epsilon}$, corresponding to a translation by $n_{\epsilon}$ sites. The equivalence
between these discrete phase shifts and spatial translations implies that the shifts have no
physical consequences, such as closing of energy gaps. Indeed, we solely observe modulation of the spectrum with periodicity $2\pi/q_n$, see Figs.~\ref{fig3} and~\ref{fig4}(c), which means that along the approximant path, $\epsilon\rightarrow 0$, the spectrum becomes independent of $\phi$.

Since the spectral projector $P_r$ [cf.~Eq.~\eqref{projector}] is composed out of the eigenstates on the model Hamiltonian, we readily obtain that $P(\phi+\epsilon)=T_{\epsilon}P(\phi)T_{\epsilon}^{-1}$ and $\partial_{\phi}P(\phi+\epsilon)= T_{\epsilon} \partial_{\phi}P(\phi) T_{\epsilon}^{-1}$. Similarly, the Chern density is periodic in $1/q_n$, $ C(\phi+\epsilon)=C(\phi)$, implying that the Chern number can be written as
\begin{align}
\int_{0}^{2\pi}d\phi\, C(\phi)  =q_n\int_{0}^{2\pi/q_n}d\phi\, C(\phi)=2\pi C(\phi=0)+O(1/q_n)\,,
\end{align}
where in the first equality we used the above discussed periodicity of the Chern density, and in the second equality we expand the integral in powers of $1/q_n$. In the limit of $q_n\rightarrow \infty$, the approximation becomes exact and the integral becomes invariant of $\phi$. In other words, the Chern number associated with a topological pump family (as a function of $\phi$), is equivalent, when the spatial modulation is quasiperiodic, to the Chern density of a single chain at a given $\phi$ times a factor of $2\pi$. Note that the dependence of $\phi$ is required for the evaluation of the Chern density, i.e., the latter can be seen as the response of the chain to a potential shift, cf.~Eq.~\eqref{pot}.

Using this argument, we can associate a 2D gap topological index (Chern number) to a gap of a 1D quasiperiodic model. In other words, using the quasiperiodic order, we quantize/flatten the Chern density such that it serves as a topological index for the 1D chain. This construction locks the Diophantine/Streda relation described above, such that at a given filling factor, when a gap occurs, an associated single-valued Chern density appears that corresponds to the Chern number of the related topological pump divided by $2\pi$. This discussion is transferable to other quasicrystals that have a Hamiltonian that is periodic with a phase shift $\phi$, which can be compensated by a translation, including in higher-dimensions~\cite{Kraus2013}. 

This topological characterization has physical implications both in the bulk and at the boundary of the quasiperiodic system: (i) moving from one quasiperiodic modulation $\tau_1$ to another $\tau_2$ will induce rearrangements of the bulk gaps (phase transitions), at energy scales corresponding to the approximant at which the two modulation separate from one another~\cite{bellissard_gap_1992}; (ii) at any periodic approximant level the topological phase transitions will be avoided with a correction of $O(1/q_n)$~\cite{Kraus2012a,Kraus2013comm,Kraus2016}; (iii) the above two points manifest at a smoothly-interpolating spatial boundary; there, subgap boundary modes with increasing density appear as a function of the boundary length~\cite{Verbin2013}; (iv) the filling-factor gap labels can appear in bulk diffraction experiments~\cite{Dareau2017}; (v) the relation to the topological pump implies that at a finite boundary to the vacuum, the bulk gap will be densely filled by subgap modes as a function of $\phi$ or equivalently, as a function of removing the last site of a half-infinite chain~\cite{Kraus2012a,Verbin2015,Kraus2016,Bloch2017,Prodan2019}, which can also be understood as a result of the mechanical work associated with the bulk Chern density, manifesting at the boundary~\cite{kellendonk2019bulk}; and (vi) at higher dimensions $D>1$, the boundary manifests a variety of effects with co-dimension up to $D$, corresponding to bulk responses of higher Chern numbers~\cite{Kraus2013,zilberberg2018photonic,petrides2018six,petrides2020higher}.

Before moving to discuss experimental realizations of the topological consequences, we now highlight once more the assumptions that we relied on. We thus discuss caveats regarding the order of limits taken in our analysis, and challenges in generalizing the results to other types of quasicrystals. Whereas labeling of gaps using filling factors is possible for various quasiperiodic models~\cite{bellissard_gap_1992}, here we discuss also another index (related to the Chern number of a higher dimensional quantum Hall state), which is associated with physical observables. This index arises as a response to a continuous potential shift $\phi$. Considering discrete quasiperiodic potentials, however, will show discontinuities that will complicate such a construction. Specifically, the interpolating potential~\eqref{pot} supports our analysis everywhere except at  $\beta=\infty$. Therefore, we must first take the approximant path to infinity before we can take the discrete step limit. Crucially, along the path, the characterization strictly relies on the fact that the gap stays open for all $\phi$ and $\theta$, and as a function of the approximant scaling, cf.~Fig.~\ref{fig5}. 

Due to these caveats, it remains to be explored how to associate Chern numbers to discrete chains such as the Thue-Morse and Period-Doubling chains, where a relation to topological boundary effects and pumping was not established. Considering cut-and-project quasicrystals, the interpolating potential~\eqref{pot} is equivalent to  ``smoothening'' the cut's acceptance window, whereas we study the response system to a phason shift of the window, cf.~Fig.~\ref{fig2}(b) and Ref.~\cite{kraus2012}. Such a smoothening in higher-dimensional quasicrystals is challenging, e.g., in the Amman-Beekner tiling the cut-window is octagonal, and can be modified in different ways. Additionally, in the presence of several quasiperiodic potentials, the length scales competition can become more elaborate than discussed here, and the gaps can close as a function $\phi$ and $\theta$.

We would like to also emphasize that we have dealt here with single-particle Hamiltonian models. The delicate length scale competition required for the topological characterization of quasicrystals can be easily perturbed. For example, disorder or dissipation  produce a self-energy broadening  that will be able to close small spectral gaps associated with long-ranged deviations from periodicity, i.e., they truncate long-ranged length scale.  Alternatively, perturbing quasicrystals with strong potentials that close spectral gaps arising from the quasiperiodic length scales competition can generate real-space curvatures and observed chiral edge modes, which are not related to the higher-dimensional topology discussed here~\cite{bandres2016topological}. 
Similarly to disorder and dissipation, many-body interactions have also been shown to act as a cut-off on the length scale at which the incommensuration prevails~\cite{vidal1999correlated,vidal2001interacting}. At the same time, new results recently appeared, where strongly-correlated quasiperiodic spin chains exhibit topological excitation gaps~\cite{Lado2019} or robust topological edge modes that are strongly shaped by the interaction~\cite{liu2020topological}. Furthermore, anomalous inverse compressibility is predicted to occur due to inter-band interactions~\cite{Kraus2014}.

\begin{figure}[t!]
\centering\includegraphics[width=\textwidth]{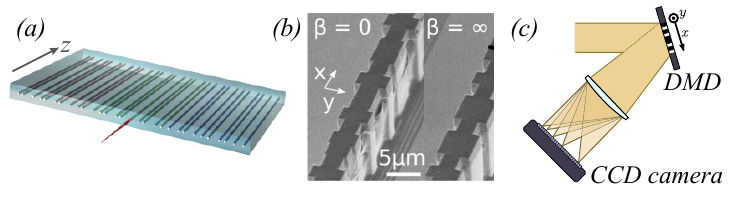}
\caption{Optical platforms for exploring the topology of quasicrystals. (a) Illustration of a coupled photonic waveguide array [cf. parallel review by Szameit], where light is injected at the input facet and emulates a tight-binding Hamiltonian evolution as it propagates through the array. Using such a platform, the dependence of the topological boundary states of quasicrystals as a function of the phason degree of freedom $\phi$ was explored~\cite{Kraus2012a,Verbin2015,zilberberg2018photonic}. Here, the red, green, and blue waveguides mark two different bulk quasiperiodic chains (red and blue) with an interpolating region (green) used to explore topological equivalence and phase transitions between different quasiperiodic models~\cite{Verbin2013}. (b) A scanning electron micrograph image of sculptured polariton semiconductor microcavities used to explore the interpolating model~\ref{pot}, taken from Ref.~\cite{strkalj2020}, cf.~Fig.~\ref{fig1}(b) and [parallel review on Exciton-polaritons for topology]. Such a setup and a similar one for acoustic waves were used to explore also the topological boundary states of the Fibonacci chain~\cite{Bloch2017,Prodan2019}. (c) Using a collimated laser beam impinging on a digital micromirror device (DMD) that encodes the Fibonacci chain spacing, the far-field diffraction pattern reveal the bulk gaps and their labels~\cite{Dareau2017}.  }\label{optics}
\end{figure}

\section{Optical experiments}
\label{sec:opt}
Contemporary technology allows for the design of structured mediums for waves, in the spirit of Fig.~\ref{fig1}(b). The chosen pattern can be thought of as a spatial modification of the effective refractive index of the medium, cf.~the variety of technological platforms covered in this special issues [cross reference here to the other review work in this issue]. Such a spatial modification generates an effective metamaterial for the waves propagating through the medium, which are then readily probed using input-output experiments. Furthermore, in the limit of deep and well-separated effective potential wells, the waves behave similarly to electrons hopping between atoms, i.e., akin to the model~\eqref{Eq:HS_1D}. 

With such platforms, both the topological bulk and boundary effects described above have been explored in 1D quasiperiodic chains, see Fig.~\ref{optics}. Specifically, using coupled photonics waveguide arrays, [cf.~review by Szameit],  topological phase transitions and topological equivalence between different quasiperiodic chains at smooth boundaries were explored, see Fig.~\ref{optics}(a) and Ref.~\cite{Verbin2013}. 
Using a similar platform, (i) the relation of 1D quasiperiodic chains to 1D topological pumps and the associated associated topological boundary states of quasiperiodic chains, were explored~\cite{Kraus2012a,Verbin2015}; (ii) The 0D topological boundary states were used to generate a high-fidelity on-chip Hong-Ou-Mandel interference of single-photons~\cite{tambasco2018quantum} [cf. contribution on quantum]; (iii) Moving to 2D topological pumps, their predicted boundary phenomena with co-dimension 1 and 2~\cite{Kraus2013}, was also observed using coupled photonic waveguide arrays~\cite{zilberberg2018photonic}, and their relationship to high-order topology was elucidated~\cite{benalcazar2017electric,petrides2020higher}. In the spirit of Fig.~\ref{fig1}(b), topological boundary modes of quasiperiodic chains were also observed in polariton semiconductor microcavities~\cite{Bloch2017} [see Fig.~\ref{optics}(b), and acoustic chambers~\cite{Prodan2019}. In diffraction experiments, bulk gap labels of the 1D Fibonacci have been observed~\cite{Dareau2017}, see Fig.~\ref{optics}(b). 

\section{Conclusions}
We have seen that the topology of quasicrystals originates from a competition between length scales. Following the approximant path, we understood that this competition ``locks'' the spectrum of the system into a very specific arrangement with well-defined integrated density of states, as well as associated bulk responses. The latter are quantized and serve for a topological classification of quasicrystals with corresponding bulk and boundary observable effects. In this sense, quasiperiodic order can be thought of as a very specific crystalline symmetry that allows for an enriched topological classification, akin to other crystalline symmetries~\cite{fu2011topological}. At the same time, the topological characterization discussed here has only been possible for certain types of quasicrystals and under specific assumptions. There is, therefore, still much to explore in this rich zoo of quasiperiodic models. Furthermore, with the advent of controllable shaping of various metamaterials for optical, acoustic, and matter waves the topological properties of quasicrystals are now observable, allowing for new discoveries and applications. 

\section*{Funding}
We acknowledge financial support from the Swiss National Science Foundation through grants PP00P2P2\_1163818 and PP00P2\_190078. 

\section*{Acknowledgments}
I would like to thank Y.~E.~Kraus, Z.~Ringel, M.~Verbin, and Y.~Lahini for starting this activity with me; J.~L.~Lado, A.~\v{S}trkalj, I.~Petrides, T.~M.~R.~Wolf, J.~Bloch, and H.~M.~Price for collaboration along this path; and E.~Prodan, J.~Kellendonk, A.~Eichler, R. Lifshits, and Y.~Avron for fruitful discussions.

\section*{Disclosures}
The author declares no conflicts of interest.

\bibliography{topQCsRev}

\end{document}